# Quantification challenges for atom probe tomography of hydrogen and deuterium in Zircaloy-4

Isabelle Mouton[a,*], Andrew J. Breen[a], Siyang Wang[b], Yanhong Chang[a], Agnieszka Szczepaniak[a], Paraskevas Kontis[a], Leigh T. Stephenson[a], Dierk Raabe[a], M. Herbig[a], T. Ben Britton[b], Baptiste Gault[a,*]

[a] Max-Planck-Institut für Eisenforschung, Max-Planck-Straße 1, 40237 Düsseldorf, Germany.
[b] Department of Materials, Royal School of Mines, Imperial College London, London, SW7 2AZ, UK

*corresponding author
*E-mail address*: i.mouton@mpie.de; b.gault@mpie.de
*Postal address:* Max-Planck-Institut für Eisenforschung, Max-Planck-Straße 1, 40237 Düsseldorf, Germany.

**Abstract:**

Analysis and understanding of the role of hydrogen in metals is a significant challenge for the future of materials science, and this is a clear objective of recent work in the atom probe tomography (APT) community. Isotopic marking by deuteration has often been proposed as the preferred route to enable quantification of hydrogen by APT. Zircaloy-4 was charged electrochemically with hydrogen and deuterium under the same conditions to form large hydrides and deuterides. Our results from a Zr hydride and a Zr deuteride highlight challenges associated to accurate quantification of hydrogen and deuterium, in particular associated to the overlap of peaks at low mass-to-charge ratio and of hydrogen/deuterium containing molecular ions. We discuss possible ways to ensure that appropriate information is extracted from APT analysis of hydrogen in a zirconium alloy systems that is important for nuclear power.

# 1   Introduction

Atom probe tomography (APT) combines a time-of-flight mass spectrometer with a projection microscope (Müller et al., 1968). The technique provides atomic scale chemical information with nanometer spatial resolution has a high sensitivity and equal efficiency across a wide range of mass-to-charge-state ratios. The main strength of APT lies in its capacity to provide a precise account of where atoms of a specific species were located within a 3D volume of probed material and provide quantitative microstructural data at the atomic scale. (Kelly & Miller, 2007; Devaraj et al., 2018).

As a mass spectroscopy technique, APT has the inherent capacity to detect light elements, including hydrogen. Yet its application to measuring hydrogen composition within metals has been hampered by the presence of hydrogen in the residual gas in the ultra-high vacuum chamber of the APT microscope. This hydrogen generates noise in the data, as the hydrogen may be adsorbed onto the tip, subsequently migrating towards the specimen apex and gets finally field desorbed and ionised (Sundell et al., 2013). This produces characteristic peaks at 1, 2 and 3 Da, corresponding respectively to $H^+$, $H_2^+$ and $H_3^+$, as well as some events uncorrelated to a pulse contributing to the general background. This "background" signal provides significant uncertainty when tracking the presence of hydrogen within the original material.

In principal the "background" hydrogen challenge can be overcome with isotopic marking, using deuterium, as the natural abundance of deuterium is very low (only 0.015%). Isotopic marking through the use of deuterium or heavy water during microstructural formation has been used to distinguish between residual hydrogen from the chamber and solute hydrogen, here deuterium, originating from the specimen itself (Gemma et al., 2011, 2007, Takahashi et al., 2010, 2018; Chen et al., 2017; Li et al., 2018).

An added complexity is that the relative amounts of the $H^+$, $H_2^+$ and $H_3^+$ peaks are highly dependent on the amplitude of the electrostatic field and on the nature of the surface (Tsong et al., 1983), making it difficult to ensure that no H gets identified as D. This problem is made more prominent as the community increasingly uses laser-pulsing, which results in operating the instrument under lower electrostatic fields (as evaporation occurs at higher temperatures (Kellogg, 1981; Vurpillot et al., 2009; Marquis & Gault, 2008)) where the relative amount of $H_2^+$ is expected to be higher (Tsong et al., 1983).

Therefore instead of looking for small concentration of trapped H solute, we investigated by APT stable bulk hydrides and deuterides in Zircalloy-4, and compare this with the hydrogen-free alloy. From a technological point of view, Zr-based alloys are amongst the materials most affected by hydrogen embrittlement when they are employed as nuclear fuel cladding in water-based nuclear reactors . Zr alloys are used in these applications as they have a low neutron absorption cross section, good corrosion resistance and appropriate mechanical strength. However, during service they corrode and pick up hydrogen and risk being affected by delayed hydride cracking (DHC) (Bair et al., 2015; Suman et al., 2015). The solubility of hydrogen in $\alpha$-Zr is less than 10 wt. ppm at room temperature (McMinn et al., 2000), and the driving force for segregation to interfaces and to form secondary phases is high. The phase diagram for the Zr-H system indicates at least four $ZrH_x$ phases at temperatures below 550 °C an hexagonal close packed (HCP) $\zeta$ with x=0.25–0.5, a face-centred tetragonal (FCT) $\gamma$ with x=1, a face-centred cubic (FCC) $\delta$ with x≈1.5–1.65 and another FCT phase $\varepsilon$ with x≈1.75-2 (Grosse et al., 2008; Li et al., 2017).

Little is known so far on the field evaporation behaviour of bulk hydrides, with only few reports on systematic studies (Takahashi et al., 2009; Chang et al., 2018). Yet, the compositional accuracy of APT is known to be strongly dependent on an appropriate set of experimental parameters, as extensively discussed for carbides, nitrides, oxides (Sha et al., 1992; Kitaguchi et al., 2014; Mancini et al., 2014; Gault, David W Saxey, et al., 2016) but also for metallic alloys (Miller, 1981; Yao et al., 2011). Voltage pulsing has typically been used for analysing solute H, in order to maximise the electrostatic field, reduce the proportion of $H_2$ and possibly mitigate surface diffusion. Here, this pulsing mode led to a premature fracture of atom probe specimens and did not entirely prevent the detection of H-containing molecular ions. We performed a systematic investigation of the results obtained from laser-pulsed APT. Following precise decomposition of the peaks in the mass-to-charge spectrum, we discuss the parameters influencing the quality of the measurement. Our results point to the need to better understand the reasons for the variations in the amount of detected residual H in comparison to solute H or D so as to enable direct analysis with no need for isotopic marking.

## 2 Experimental

Experiments were conducted on a rolled and recrystallized plate of a commercial Zircaloy-4 with a composition of Zr-1.5Sn-0.2Fe-0.1Cr wt.%. . The alloy was heat treated at 800 °C for two weeks to form large 'blocky-alpha' grains with a typical grain size larger than 200 μm, as described in Tong and Britton (Tong & Britton, 2017). Samples were electrochemically charged with hydrogen or deuterium using galvanostatic charging at a current density of 2 kA/m$^2$, using a solution of 1.5 wt. % H/D$_2$SO$_4$ in H/D$_2$O at 65 °C for 24 hours. An approximately 20 μm thick hydride layer was formed at the surface. The H/D was then redistributed inside the bulk via annealing at 400 °C for 5 hours, which was followed by furnace cooling at a rate of 0.5 °C/min. This process led to the formation of hydrides at grain boundaries and within the blocky grains, as indicated in the polarised light optical micrograph in Figure 1 (a).

The microstructure was then inspected using electron backscattered diffraction (EBSD) in a Auriga scanning electron microscope (SEM) equipped with a field emission gun (Carl Zeiss Microscopy) run at 30 kV and a probe current of 10.5 nA, using a Bruker eFlashHR detector and the Esprit 2.1 software. APT specimens were prepared using an FEI Helios dual-beam plasma focused ion beam (PFIB), following the procedure proposed by Thompson et al. (Thompson et al., 2007).

An inverse pole figure (IPF) map rendered with respect to the surface normal is shown for a typical intergranular hydride formed at a twin boundary in Figure 1(b). The misorientation angle is approximately 85° between the parent and twin α-Zr grains, which indicates that they are $\{10\bar{1}2\} <1\bar{0}11>$ extension twins, for which the theoretical misorientation angle about c-axis is ~85.22°. An orientation gradient can be easily identified within the pink hydride packet in the IPF implying misfit strain between the matrix and the δ hydrides.

The preparation protocol is summarised in Figure 1(c–e), with the hydride clearly visible both in the secondary electron micrograph of the lifted out bar sliced onto the support in Figure 1(d) and of the final specimen in Figure 1(e). The ruggedness of the interface is evident from the side views. APT results were obtained on a CAMECA local electrode atom probe (LEAP) 5000 XR. UV laser pulsing was thus used, which drastically improved yield. A base temperature of 60 K, laser pulse energy of 60 pJ, pulse repetition rate of 250 kHz and a detection rate of 1-2 ions per 100 pulses were used during the experiments.

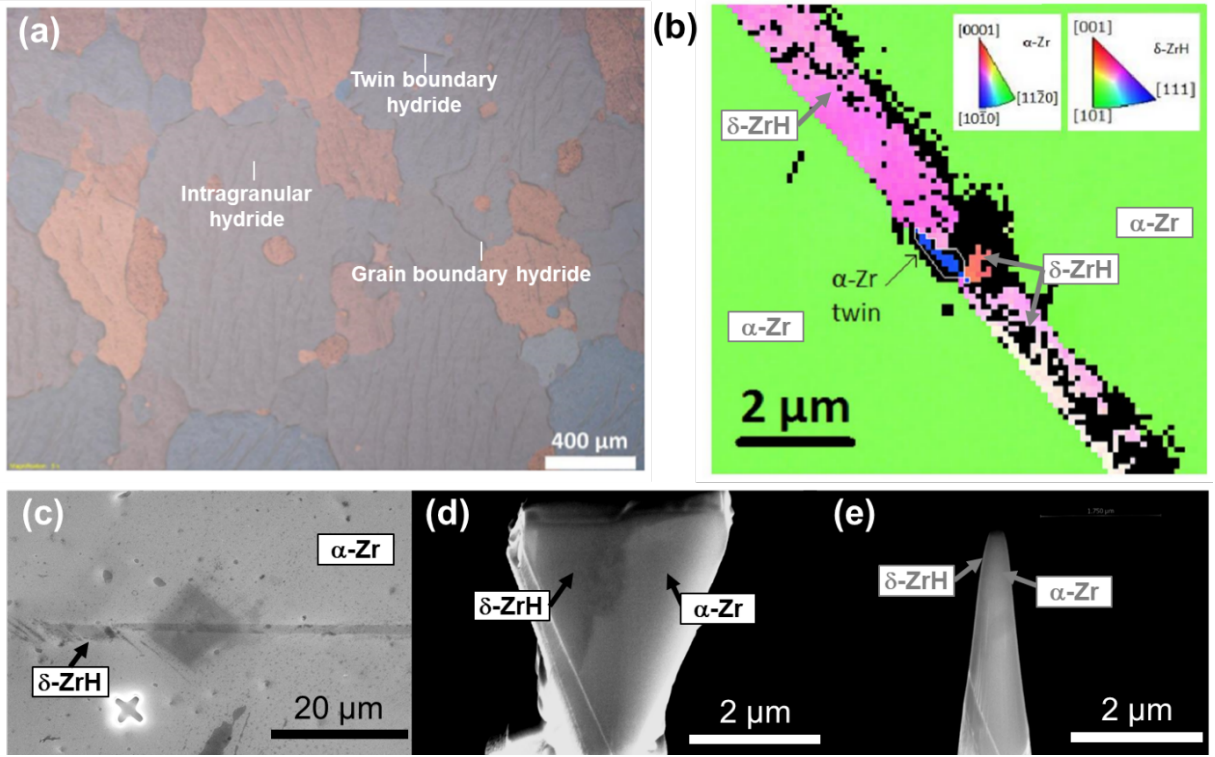

*Figure 1: (a) polarized light optical micrograph of the hydride microstructure showing both intragranular and intergranular hydrides. (b) IPF map of the selected twin boundary hydride, rendered with respect to the surface normal. Secondary electron micrographs (c) of the marked hydride to ensure that this specific hydride was analysed; (d) of a slice of the 'Toblerone-shaped' bar containing the hydride lifted-out from the material and mounted on the support; (e) final specimen. The dark contrast reveals the presence of the hydride in the final specimen.*

## 3 Methods

Based on the natural abundances of the different elements it is possible to perform a decomposition of the mass peaks. This alleviates quantification issues associated to the overlap of peaks corresponding to different ionic species that have the same mass-to-charge ratio (Miller, 2000). We implemented an automated peak decomposition procedure, akin to what was proposed in ref. (London et al., 2017), but here this procedure was either applied to the entire reconstruction or to individual bins of a composition profile. The background level due to hydrogen evaporating between pulses is subtracted prior to performing this procedure. Typical results are displayed in the form of a histogram as shown in Figure 2, where the number of ions $^mN$ in the peak with mass/charge of $m$Da is determinate ($^{45}N$ in the peak at 45Da). Each peak is a combination of the natural abundance of the Zr ($A_{Zr}$) and the proportion of $Zr^{2+}$ $ZrH^{2+}$ and $ZrH_2^{2+}$ which ca be written:

$$\begin{bmatrix} {}^{45}N \\ {}^{45.5}N \\ {}^{46}N \\ {}^{46.5}N \\ {}^{47}N \\ {}^{47.5}N \\ {}^{48}N \\ {}^{48.5}N \\ {}^{49}N \end{bmatrix} = \begin{bmatrix} A_{90_{Zr}} & 0 & 0 \\ A_{91_{Zr}} & A_{90_{Zr}} & 0 \\ A_{92_{Zr}} & A_{91_{Zr}} & A_{90_{Zr}} \\ A_{93_{Zr}} & A_{92_{Zr}} & A_{91_{Zr}} \\ A_{94_{Zr}} & A_{93_{Zr}} & A_{92_{Zr}} \\ A_{95_{Zr}} & A_{94_{Zr}} & A_{93_{Zr}} \\ A_{96_{Zr}} & A_{95_{Zr}} & A_{94_{Zr}} \\ 0 & A_{96_{Zr}} & A_{95_{Zr}} \\ 0 & 0 & A_{96_{Zr}} \end{bmatrix} \begin{bmatrix} N_{Zr^{2+}} \\ N_{ZrH^{2+}} \\ N_{ZrH_2^{2+}} \end{bmatrix}. \qquad \text{(eq. 1)}$$

As this is an overdetermined system of linear equations, a least squares solution was numerically found. Explicitly, we minimised the sum of squared residuals between the number of ions of $Zr^{2+}$, $ZrH^{2+}$, $ZrH_2^{2+}$ determined experimentally and the numerically by trial. The precision of the decomposition depends on the residual number of ions after decomposition. The resulting decomposition is presented as coloured bars which represent the relative amplitude of each of the identified species. The histogram in Figure 2 was obtained from an uncharged sample, and yet some $ZrH^{2+}$ can be seen, whereas no peak of $ZrH_2^{2+}$ or $ZrD^{2+}$ can be measured.

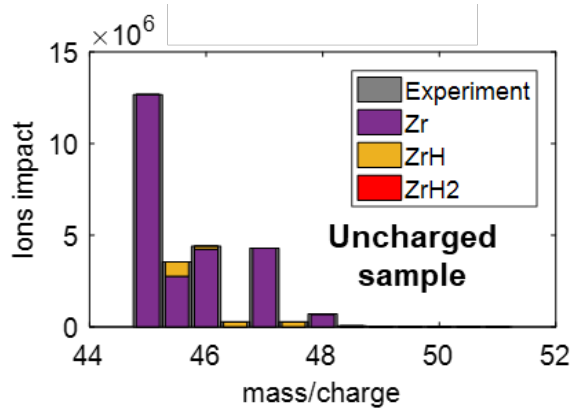

*Figure 2: typical histogram resulting from the decomposition of the mass peaks based on the elements' natural abundances.*

## 4 Results and discussion

### 4.1 Species distribution

Thin slices (10nm) through the tomographic reconstructions from representative APT datasets for the metal-hydride and metal-deuteride interfaces are shown in Figure 3(a) and (b). This allows for direct comparison of the detection of H and D within a stable phase and as solute in the metallic matrix. Similar features can be observed in both, with a segregation of Sn at a very rough interface, as reported in a separate article (Breen & et al., n.d.). This segregation at the growth front of the hydride into the metal was recently reported. Sections of the mass spectrum obtained for these two datasets are displayed in Figure 3 (c) and (d). For the hydride (in black), peaks appear at 1 and 2 Da, corresponding to $H^+$ and $H_2^+$ as expected. For the deuteride (in red), the amplitude of the peak at 2 Da corresponding most likely to $D^+$ is higher relative to the $H^+$ peak. This peak likely also contains a certain amount of $H_2^+$ which is undetermined. Additional peaks at 3 and 4 Da corresponding to $HD^+$ and $D_2^+$ are also apparent. As evidence by the mass spectrum of the H-charged sample, the presence of triple ions ($H_3^+$ and so $DH_2^+$) are negligible. The possible combinations of molecular ions in this range of mass-to-charge ratios are listed in Table 1.

| Mass-to-charge (Da) | Combinations of possible species | Uncharged or H-charged | D-charged |
|---|---|---|---|
| 1 | $H^+$ | 1 H | 1 H |
| 2 | $H_2^+ + D^+$ | 2 H | 1 D |
| 3 | $H_3^+ + DH^+$ | 3 H | 1 D +1H |
| 4 | $H_4^+ + DH_2^+ + D_2^+$ |  | 2 D |

Table 1: Likely combinations of possible species detected at in the range of 1–4 Da for the hydrided and deuterated samples.

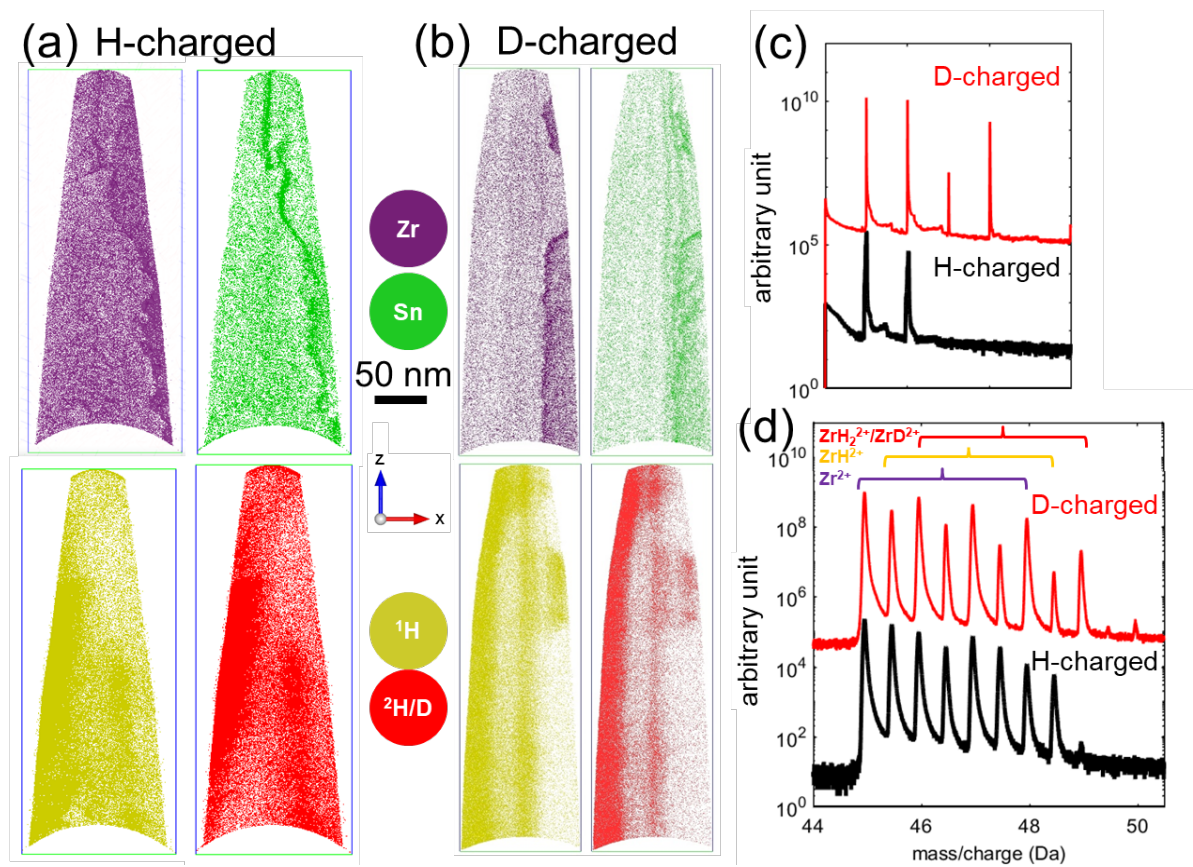

Figure 3: Distribution of species for Zr, Sn, $^1H$ and $^2H$ at the (a) hydride/α-Zr interface and (b) deuteride/α-Zr interface. Sections of the mass spectrum in the range of (c) 1–6 Da and (d) 44–50.5 Da obtained for the hydride (black) and the deuteride (red).

Peaks corresponding to H- and D- containing molecular ions, e.g. $ZrH^{2+}/ZrD^{2+}$ are also present in both datasets, as seen in Figure 3(d). The $ZrH^{2+}/ZrD^{2+}$ are present in the region of the dataset corresponding to the H/D-rich phase but surprisingly also within the α-Zr phase. This is evidenced in Figure 4 that shows the results from the peak decomposition procedure described above applied to regions of interest extracted from the α-Zr and the H/D-rich phase in datasets of both the hydrogenated and deuterated samples. In the H-charged specimen (α matrix phase in Fig.4 a and H-rich phase in Fig. 4c), peaks associated to Zr and ZrH are measured, but none to $ZrH_2$. Whereas in the D-charged specimen (α matrix phase in Fig.4 b and D-rich phase in Fig. 4d), the same peaks appear but high peaks labelled as $ZrH_2$, likely corresponding mostly to ZrD, are also present. The difference in the relative amplitude of the peaks from the different species is evident between the two samples. Interestingly, the α-Zr matrix consistently contains a relative high level of H or D.

## 4.2 Compositional analysis

To accurately measure the composition of H/D in the different phases represent two major challenges. First, the amount of H originating from residual gases and/or the specimen needs to be established and, second, the possible overlap between $H_2$ and D needs to be disentangled.

The presence of a significant amount of hydrogen in the analysis of both the uncharged and D-charged sample may first be related to:

- Residual H from within the APT chamber that can be field ionised or field desorbed from the specimen in the form of atomic, homomolecular or heteromolecular ions (Sundell et al., 2013).
- Partial hydration during deuteration or the subsequent exchange of D by H as the specimen was kept in air for several weeks after preparation.
- A likely source of H is the preparation of the specimens themselves. Samples have been mechanically polished using water-based solution which is known to cause hydride formation (Kestel, 1986). This is also known to occur via FIB-based specimen preparation of Zr and Ti (Shen et al., 2016; Chang et al., 2018).

A combination of these effects likely explains the $ZrH^{2+}$ peaks, as well as the H peaks (not shown here), observed for data obtained respectively from an uncharged sample (Figure 2) and from a D-charged sample (Figure 4 (b) and (d)).

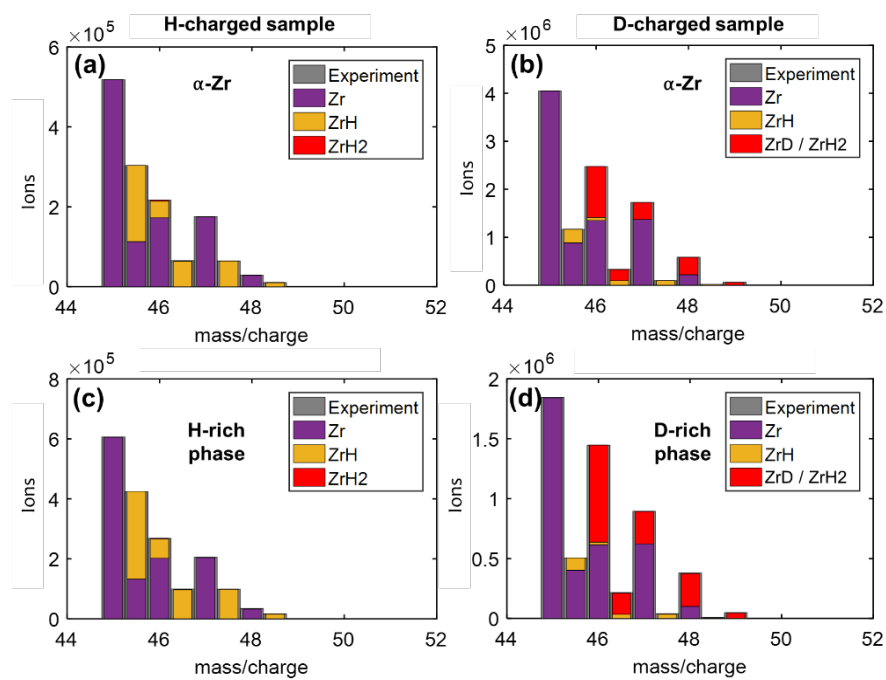

Figure 4: Results of the peak decomposition for the $Zr^{2+}$ peaks in the α-Zr in (a) the hydrided and (b) the deuterated samples. Similar results $Zr^{2+}$ peaks in (a) the hydride and (b) the deuteride.

H quantification in α and H/D-Zr phase can be obtained from composition profiles along rectangular regions-of-interest positioned perpendicularly to the interface. Because of the roughness of the interface, readily visible in Figure 3, only thin slices with long sections of parallel to the interface were considered, the precise size is 15 x 25 x 75 nm$^3$ and 15 x 60 x 100 nm$^3$ for H and D charged sample). Figure 5 (a) and (b) represented the composition profiles across the hydride/α-Zr and deuteride/α-Zr interface for respectively Zr and H after application of the complete peak decomposition protocol.

The results in Table 1 confirm that there is significant uncertainty regarding the possible overlap between $H_2$ and D, and its amplitude. Therefore, to quantify H/D, the different peaks are indexed as in the Table 1 (two last columns) for H/D-charged sample. For the H-charged sample, the H peak was quantify as coming all from the sample, ignoring the residual H from the ultra-high vacuum chamber. Therefore, the H composition in both α matrix and hydride phases will be slightly overestimated. For the D-charged sample, the peak at 2Da was considered as being exclusively D. The absolute composition of D reported here is hence likely overestimated.

A progressive drop in the level of H is observed in the α-Zr that could be related to fluctuations related to variations in the local amplitude of the electrostatic field necessary to cause field evaporation when evaporating the deuteride compared to evaporating the α-Zr. In both cases, the level of H or D within the stable hydride or deuteride phases is in the range of 60–65 at%. The level of H or D in the α-Zr matrix is surprisingly high, in the range of approximatively 18–22 at%, which is well above the expected maximum solubility of approximatively 2 at% (Giroldi et al., 2009). Contrary to the uncharged Zr sample (Figure 2), the high content of H/D in the α-Zr for H/D charged sample is mainly arising from decomposition of ZrH and ZrD, show Figure 5 (b) and (d) respectively. Therefore, H/D atoms are likely introduced during the charging process, but presence of H/D may also be related to a different maximal solubility in electrochemical-charging conditions. The chemical potential of the H/D in the solution is rather different to that in the gaseous environment that is usually employed to perform charging and solubility measurements (Yamanaka et al., 1995). The presence of defects near the polished surface or strain caused by the growth of the hydride in the matrix near the precipitate can also influence the solubility of H (Christensen et al., 2015).

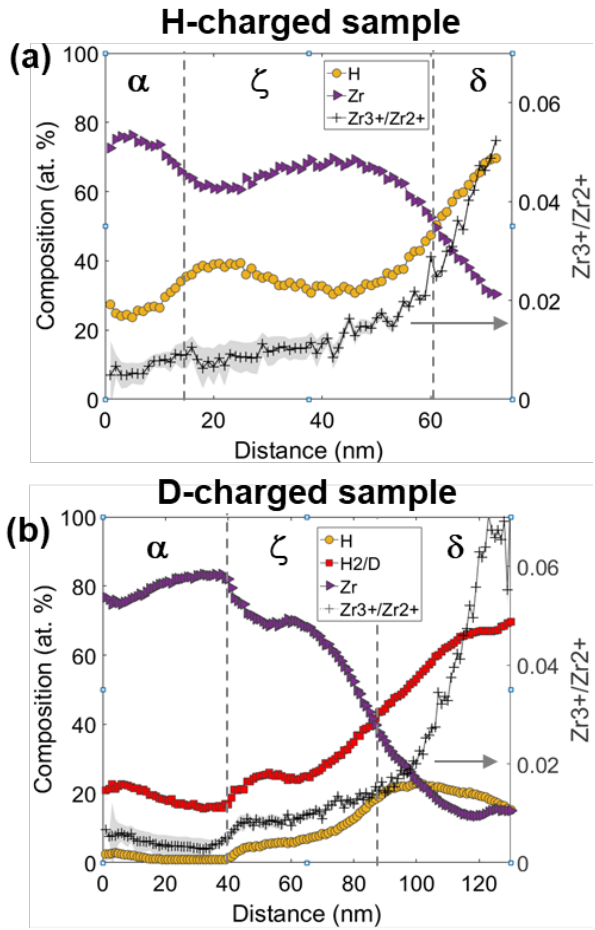

*Figure 5: Profile chemical analysis across (a) hydride/α-Zr interface and (b) deuteride/α-Zr interface*

### 4.3 Electrostatic field dependence

The evolution of the relative charge-states of a single element can be used as a tracer to investigate the local variations of the electrostatic field within a single dataset (Kingham, 1982; Müller et al., 2011; Shariq et al., 2009). Here, the relative amplitude of $Zr^{3+}$ to $Zr^{2+}$ across the interface was measured. In Figure 6, the composition of H- or D-containing species is plotted against this ratio. A higher electrostatic field is observed within the hydride and deuteride, as compared to the matrix. In the composition profile obtained from the H-charged sample shown in Figure 6 (a), the progressive drop in amplitude of ZrH as the electrostatic field increases likely indicates a field-assisted dissociation of this molecular ion (Tsong et al., 1983; Gault, D.W. David W Saxey, et al., 2016). A similar behaviour is observed in the D-charged sample composition profile for the peak labelled $ZrH_2$, likely to be almost exclusively ZrD, in Figure 6 (b). Interestingly, for the case of the deuteride, the composition of H (at 1Da) progressively drops as the field increases.

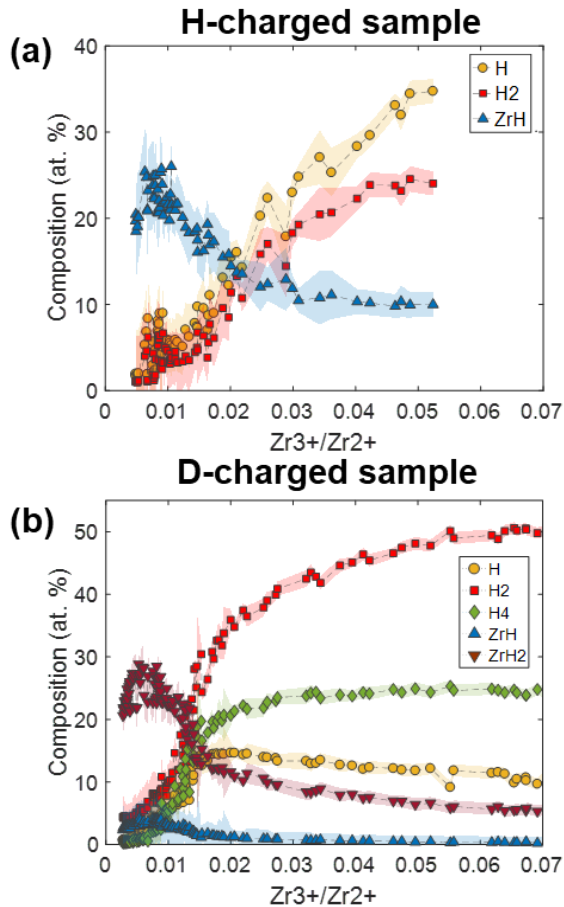

*Figure 6: Composition of the various (a) H- or (b) D-containing species dependence with the local electrostatic field.*

### 4.4 General considerations

The analyses presented showcase the difficulties inherent to quantifying hydrogen within the microstructure of a metallic alloys, even when there is a stable hydride-rich phase present.

First, it is not straightforward to separate the overlap of $H_2$ and D, and to a lesser extent, $H_3$ and DH. The strong dependence on the electrostatic field of the atomic and the molecular nature of the ions detected makes difficult (if not impossible) to distinguish residual H from solute H. Earlier work, particularly by Tsong and co-workers, highlighted the diversity of parameters that contribute to fluctuations in the relative composition of $H^+$, $H^{2+}$ and $H^{3+}$, including the nature of the surface that is likely to have a catalytic effect on the formation of $H_3$ or the splitting of $H_2$ for instance (Ai & Tsong, 1984; Chi-fong Ai & Tsong, 1984). The relative amounts of $H_2$ and $H_3$ are highly dependent on the local magnitude of the electrostatic field that can fluctuate significantly across the field-of-view and over time during an experiment. This is clearly demonstrated in Figure 6. These problems can be made more prominent by the use of laser pulsing that often result in more pronounced variations of the electrostatic field across the field of view (Gault et al., 2010) and field evaporation at higher temperature.

Performing a local analysis is possible, but reducing the sample size reinforces the relative importance of statistical fluctuations, making the measurement potentially significantly less precise. Using D-charging as a replacement for H is good solution to locate the H ions in the microstructure (provided that physical differences between H and D during microstructure formation can be assumed to be negligible). Nevertheless, dependent of the material and the experimental condition (cryo preparation, analysis with laser pulse), D charging can increases the complexity of quantification. Thus, although isotopic marking by deuteration is elegant and attractive, it is not a miraculous solution. This supports previous work Sundell et al. (Sundell et al., 2013). Indeed, a significant fraction of the H/D is detected as part of molecular ions, and their respective proportions are highly dependent on the amplitude of the electrostatic field.

A preferred route would be to gain a better understanding of how residual and solute hydrogen behave during the course of the analysis, thereby potentially enabling to distinguish between the two with no need for deuteration. The supply of the residual gas to the ionisation or field desorption zone is mostly related to the diffusion of adsorbed hydrogen on the surface of the specimen (Sundell et al., 2013). However, the detailed mechanisms are still poorly understood in the case of very low partial pressures of gas and when multiple phases intersect the surface (Orloff, 2008). Overall, our results agree with reports on other materials which find maximising the electrostatic field will tend to reduce the influence of the residual hydrogen both in the form of molecular hydrogen and H-containing molecular ions. There might also be a possibility to find appropriate descriptors and track their evolution locally during the experiment. The electrostatic field is clearly one such descriptors, which was discussed at length herein, but others may exist. The deployment of machine learning techniques to aid in separation could later lead to interesting findings in this area.

# 5  Conclusions

In summary, we have characterised by APT stable Zr-hydrides and deuterides, and their interface with the α-Zr matrix, in electrochemically charged Zircalloy-4. We highlighted the dependence of the detected atomic or molecular ionic species detected on the local strength of the electrostatic field at the specimen's surface. The influence of the residual H from the ultra-high vacuum chamber was shown to become less pronounced as the electrostatic field increases, which hints towards that the field condition should be maximising for H quantification. Finally, we discussed isotopic marking which helps with assessing the locations of high concentration of hydrogen within the microstructure, but adds to the complexity associated to quantifying H by APT.

## Acknowledgements

Uwe Tezins and Andreas Sturm are thanked for their support to the FIB and APT facilities at MPIE. AB, MH, DR, BG acknowledge the Deutsche Forschungsgemeinschaft (DFG) for partially funding this research through SFB 761 'Stahl ab initio'. AB also acknowledges the Alexander von Humboldt Foundation (AvH) for partially funding this research. The authors are grateful for the Max-Planck Society and the BMBF for the funding of the Laplace and the UGSLIT projects respectively. TBB thanks the Royal Academy of Engineering for support of his Research Fellowship. TBB and SW acknowledge support from the HexMat programme grant (EP/K034332/1).